\documentstyle[12pt]{article}
\begin{document}

\def\theequation {\thesection.\arabic{equation}}
\makeatletter\@addtoreset {equation}{section}\makeatother

\title{\bf Spectral decomposition for the Dirac system 
associated to the DSII equation}

\author{Dmitry E. Pelinovsky and Catherine Sulem \\
\normalsize Department of Mathematics, University of Toronto \\
\normalsize Toronto, Ontario, Canada, M5S 3G3}

\date{}
\maketitle\thispagestyle{empty}

\begin{abstract}
A new (scalar) spectral decomposition is found for the Dirac system 
in two dimensions associated to the focusing Davey--Stewartson II 
(DSII) equation. Discrete spectrum in the spectral problem corresponds 
to eigenvalues embedded into a two-dimensional essential spectrum. We 
show that these embedded eigenvalues are structurally unstable under 
small variations of the initial data. This instability leads to the 
decay of localized initial data into continuous wave packets prescribed 
by the nonlinear dynamics of the DSII equation.
\end{abstract}

\vspace{1cm}
\begin{center}
submitted to  \\
{\bf Inverse Problems}
\end{center}

\newpage

\section{Introduction}

Gravity-capillary surface wave packets are described by the 
Davey--Stewartson (DS) system \cite{DS} which is integrable by 
inverse scattering tranform in the limit of shallow water 
\cite{AC}. In this paper, we study the focusing DSII equation 
which can be written in a complex form,
\begin{eqnarray}
\nonumber i u_t + u_{zz} + u_{\bar{z} \bar{z}} + 
4 ( g + \bar{g} ) u & = & 0, \\
\label{compl} 2 g_{\bar{z}} - \left( |u|^2 \right)_{z} & = & 0,  
\end{eqnarray}
where $z = x + i y$, $\bar{z} = x - i y$, $u(z,\bar{z},t)$ and 
$g(z,\bar{z},t)$ are complex functions. This equation appears 
as the compatibility condition for the two-dimensional Dirac system, 
\begin{equation}
\label{Dir}
\varphi_{1 \bar{z}} = - u \varphi_2, \;\;\;\;
\varphi_{2 z} = \bar{u} \varphi_1, 
\end{equation}
coupled to the equations for the time evolution of the eigenfunctions, 
\begin{eqnarray}
\nonumber i \varphi_{1 t} + \varphi_{1 zz} + u \varphi_{2 \bar{z}} 
- u_{\bar{z}} \varphi_2 + 4 g \varphi_1 & = & 0, \\ \label{temp}
- i \varphi_{2 t} + \varphi_{2 \bar{z} \bar{z}} + \bar{u}_z 
\varphi_1 - \bar{u} \varphi_{1 z} + 4 \bar{g} \varphi_2 & = & 0. 
\end{eqnarray}

The DSII equation was solved formally through the $\bar{\partial}$ 
problem of complex analysis by Fokas and Ablowitz \cite{FA} and 
Beals and Coifman \cite{BC}. Rigorous results on existence and 
uniqueness of solutions of the initial-value problem were established 
under a small-norm assumption \cite{FS}. The small-norm assumption 
was used to eliminate homogeneous solutions of equations of the 
inverse scattering which correspond to bound states and radially 
symmetric localized waves (lumps) of the DSII equation. When the 
potential in the linear system becomes weakly localized (in $L^2$ 
but not in $L^1$), homogeneous solutions may exist and the analysis 
developed in Ref. \cite{FS} is not applicable. 

The lump solutions were included formally in \cite{FA}, where their 
weak decay rate was found, $u \sim {\rm O}(R^{-1})$ as $R = 
\sqrt{x^2 + y^2} \to \infty$. This result is only valid for 
complexified solutions of the DSII equation (when $|u|^2$ is 
considered to be complex). The reality conditions were incorporated 
in the work of Arkadiev et al. \cite{APP} where lumps were 
shown to decay like $u \sim {\rm O}(R^{-2})$. Multi-lump solutions 
were expressed as a ratio of two determinants \cite{APP}, or, in 
a special case, as a ratio of two polynomials \cite{APP2} but their 
dynamical role was left out of consideration.

Recently, structural instability of a single lump of the DSII equation 
was reported by Gadyl'shin and Kiselev \cite{GK1,GK2}. The authors used 
methods of perturbation theory based on completeness of {\em squared} 
eigenfunctions of the Dirac system \cite{K1,K2}. A similar conclusion 
was announced by Yurov who studied Darboux transformation of the 
Dirac system \cite{Yu1,Yu2}. 

In this paper, we present an alternative solution of the problem 
of stability of multi-lump solutions of the DSII equation. The approach 
generalizes our recent work on spectral decomposition of a linear 
time-dependent Schr\"{o}dinger equation with weakly localized 
(not in $L^1$) potentials \cite{PS}. We find a new spectral 
decomposition in terms of {\em single} eigenfunctions of the 
Dirac system. Surprisely enough, the two-component Dirac system in 
two dimensions has a scalar spectral decomposition. In contrast,  
we recall that the Dirac system in one dimension (the so-called AKNS 
system) has a well-known 2 x 2 matrix spectral decomposition \cite{AKNS}. 

Using the scalar spectral decomposition, we associate the multi-lump 
potentials with eigenvalues embedded into a two-dimensional essential 
spectrum of the Dirac system. Eigenvalues embedded into a 
one-dimensional essential spectrum occur for instance for the 
time-dependent Schr\"{o}dinger problem \cite{PS,SW}. They were found 
to be structurally unstable under a small variation of the potential. 
Depending on the sign of the variation, they either disappear or 
become resonant poles in the complex spectral plane which 
correspond to lump solutions of the KPI equation \cite{PS}.

For the Dirac system in two dimensions, the multi-lump potentials and 
embedded eigenvalues are more exotic. The discrete spectrum of the 
Dirac system is separated from the continuous spectrum contribution 
in the sense that the spectral data satisfy certain 
constraints near the embedded eigenvalues. These constraints are 
met for special solutions of the DSII equation such as lumps, but 
may not be satisfied for a generic combination of lumps and radiative 
waves. As a result, embedded eigenvalues of the Dirac system generally 
disappear under a local disturbance of the initial data. Physically, 
this implies that a localized initial data of the DSII equation decays 
into radiation except for the cases where the data reduce to special 
solutions such as lumps.

The paper is organized as follows. Elements of inverse scattering 
for the Dirac system are reviewed in Section 2, where we find that 
the discrete spectrum of the Dirac system is prescribed by certain 
constraints on the spectral data. Spectral decomposition is described 
in Section 3 with the proof of orthogonality and completeness relations 
through a proper adjoint problem. The perturbation theory for lumps 
is developed in Section 4 where some of previous results \cite{GK1,GK2} 
are recovered. Section 5 contains concluding remarks. Appendix A 
provides a summary of formulas of the complex $\bar{\partial}$-analysis 
used in proofs of Section 3.

\section{Spectral Data and Inverse Scattering}

Here we review some results on the Dirac system (\ref{Dir}) and 
discard henceforth the time dependence of $u$, $g$ and 
\mbox{\boldmath $\varphi$}. The potential $u(z,\bar{z})$ is 
assumed to be non-integrable ($u \notin L^1$) with the boundary 
conditions, $u \sim {\rm O}(|z|^{-2})$ as $|z| \to \infty$. 

\subsection{Essential Spectrum of the Dirac System}

We define the fundamental matrix solution of Eq. (\ref{Dir}) 
in the form \cite{APP}, 
\begin{equation} \label{fund}
\mbox{\boldmath $\varphi$} = \left[ \mbox{\boldmath $\mu$} 
(z,\bar{z},k,\bar{k}) e^{ikz}, \;\;\; \mbox{\boldmath $\chi$} 
(z,\bar{z},k,\bar{k}) e^{- i \bar{k}\bar{z}} \right],
\end{equation}
where $k$ is a spectral parameter, 
$\mbox{\boldmath $\mu$}(z,\bar{z},k,\bar{k})$ and 
$\mbox{\boldmath $\chi$}(z,\bar{z},k,\bar{k})$ satisfy the  system,
\begin{eqnarray}
\label{Dir1}
\mu_{1 \bar{z}} & = & - u \mu_2, \;\;\;\;\;\;\;\;\;\;\;\;
\mu_{2 z} = - i k \mu_2 + \bar{u} \mu_1, \\
\label{Dir2}
\chi_{1 \bar{z}} & = & i \bar{k} \chi_1 - u \chi_2, \;\;\;\; 
\chi_{2 z} = \bar{u} \chi_1.
\end{eqnarray}
It follows from Eqs. (\ref{Dir1}) and (\ref{Dir2}) that \mbox{\boldmath 
$\mu$} and \mbox{\boldmath $\chi$} are related by the symmetry constraint, 
\begin{equation}
\label{chi}
\mbox{\boldmath $\chi$}(z,\bar{z},k,\bar{k}) = \mbox{\boldmath $\sigma$} 
\mbox{\boldmath $\bar{\mu}$}(z,\bar{z},k,\bar{k}), \;\;\;\;\;\;\;
\mbox{\boldmath $\sigma$} = \left( \begin{array}{cc} 0 & -1 \\ 
1 & 0 \end{array} \right).
\end{equation}
We impose the boundary conditions for $\mbox{\boldmath 
$\mu$}(z,\bar{z},k,\bar{k})$,
\begin{equation}
\label{limit}
\lim_{|k| \to \infty} \mbox{\boldmath $\mu$}(z,\bar{z},k,\bar{k}) 
= {\bf e}_1 = \left( \begin{array}{cc} 1 \\ 0 \end{array} \right).
\end{equation}
Solutions of Eq. (\ref{Dir1}) with boundary conditions (\ref{limit}) 
can be expressed through the Green's functions as Fredholm's 
inhomogeneous integral equations \cite{FA,BC}, 
\begin{eqnarray}
\label{mu1} 
\mu_1(z,\bar{z},k,\bar{k}) & = & 1 - \frac{1}{2 \pi i} \int\!\!\int 
\frac{d z' \wedge d \bar{z}'}{z' - z} (u \mu_2 )(z',\bar{z}'), \\ 
\label{mu2} \mu_2(z,\bar{z},k,\bar{k}) & = & \frac{1}{2 \pi i} \int\!\!\int 
\frac{d z' \wedge d \bar{z}'}{\bar{z}' - \bar{z}} 
(\bar{u} \mu_1 )(z',\bar{z}') e^{- i k(z-z') - 
i \bar{k} (\bar{z} - \bar{z}')}.
\end{eqnarray}

Values of $k$ for which the homogeneous system associated to Eqs. 
(\ref{mu1}) and (\ref{mu2}) has bounded solutions are called 
{\em eigenvalues} of the discrete spectrum of the Dirac system. 
Let us suppose  that the homogeneous solutions (eigenvalues) are not 
supported by the potential $u(z,\bar{z})$. We evaluate the departure 
from analyticity of \mbox{\boldmath $\mu$} in the $k$ plane by 
calculating the  derivative $\partial \mbox{\boldmath $\mu$} / 
\partial \bar{k}$ directly from the system (\ref{mu1})-(\ref{mu2}) 
\cite{FA} as  
\begin{equation}
\label{d-bar}
\frac{\partial \mbox{\boldmath $\mu$}}{\partial \bar{k}} = 
b(k,\bar{k}) {\bf N}_{\mu} (z,\bar{z},k,\bar{k}).
\end{equation}
Here $b(k,\bar{k})$ is the spectral data,
\begin{equation}
\label{b}
b(k,\bar{k}) = \frac{1}{2 \pi} \int\!\!\int dz \wedge d \bar{z} 
\left( \bar{u} \mu_1 \right)(z,\bar{z}) e^{i(kz + \bar{k} \bar{z})},
\end{equation}
and ${\bf N}_{\mu}(z,\bar{z},k,\bar{k})$ is a solution of Eq. (\ref{Dir1}) 
which is linearly independent of $\mbox{\boldmath $\mu$}(z,\bar{z},
k,\bar{k})$ and satisfies the boundary condition, 
\begin{equation}
\label{N_m}
\lim_{|k| \to \infty} {\bf N}_{\mu}(z,\bar{z},k,\bar{k}) 
e^{i(kz + \bar{k} \bar{z})} = {\bf e}_2 = 
\left( \begin{array}{cc} 0 \\ 1 \end{array} \right). 
\end{equation}
This solution can be expressed through the Fredholm's inhomogeneous 
equations,
\begin{eqnarray}
\label{Nmu1} 
N_{1 \mu}(z,\bar{z},k,\bar{k}) & = & \frac{1}{2 \pi i} \int\!\!\int 
\frac{d z' \wedge d \bar{z}'}{z' - z} (u N_{2 \mu} )(z',\bar{z}'), \\ 
\label{Nmu2} N_{2 \mu}(z,\bar{z},k,\bar{k}) & = & e^{-i(kz+\bar{k}\bar{z})} 
+ \frac{1}{2 \pi i} \int\!\!\int \frac{d z' \wedge d \bar{z}'}{\bar{z}' 
- \bar{z}} (\bar{u} N_{1 \mu} )(z',\bar{z}') e^{- i k(z-z') - 
i \bar{k} (\bar{z} - \bar{z}')}.
\end{eqnarray}
The following reduction formula \cite{FA} connects ${\bf N}_{\mu}
(z,\bar{z},k,\bar{k})$ and \mbox{\boldmath $\mu$}$(z,\bar{z},k,\bar{k})$, 
\begin{equation}
\label{N_2}
{\bf N}_{\mu}(z,\bar{z},k,\bar{k}) = \mbox{\boldmath 
$\chi$}(z,\bar{z},k,\bar{k}) e^{-i(kz+\bar{k} \bar{z})} 
= \mbox{\boldmath $\sigma$} \bar{\mbox{\boldmath 
$\mu$}}(z,\bar{z},k,\bar{k}) e^{-i(kz + \bar{k} \bar{z})}.
\end{equation}

If the potential $u(z,\bar{z})$ has the boundary values $u \sim 
{\rm O}(|z|^{-2})$ as $|z| \to \infty$, ($u \notin L^1$), then, 
the integral kernel in Eq. (\ref{b}) is not absolutely integrable, 
while Eqs. (\ref{mu1}) and (\ref{mu2}) are still well-defined. We 
specify complex integration in the $z$-plane of a non-absolutely 
integrable function $f(z,\bar{z})$ according to the formula,
\begin{equation}
\label{comp_integr}
\int\!\!\int dz \wedge d\bar{z} f(z,\bar{z}) = \lim_{R \to \infty} 
\int\!\!\int_{|z| \leq R} dz \wedge d \bar{z} f(z,\bar{z}).
\end{equation}
The same formula is valid for integrating eigenfunctions of the Dirac 
system in the $k$ plane as well. In Section 3, we use (\ref{comp_integr}) 
when computing the inner products and completeness relations for the 
Dirac system and its adjoint.

\subsection{The Discrete Spectrum}

Suppose here that integral equations (\ref{mu1}) and (\ref{mu2}) have
homogeneous solutions at an eigenvalue $k = k_j$. The discrete spectrum 
associated to multi-lump potentials was introduced in Refs. \cite{FA,APP}. 
Here we review their approach and give a new result (Proposition 2.1) 
which clarifies the role of discrete spectrum in the spectral problem 
(\ref{Dir1}). 

For the discrete spectrum associated to the multi-lump potentials, an 
isolated eigenvalue $k = k_j$ has double multiplicity with the 
corresponding two bound states ${\bf \Phi}_j(z,\bar{z})$ and 
${\bf \Phi}'_j(z,\bar{z})$ \cite{APP}. The bound state 
${\bf \Phi}_j(z,\bar{z})$ is a solution of the homogeneous equations, 
\begin{eqnarray}
\label{hom1} 
\Phi_{1j}(z,\bar{z}) & = & - \frac{1}{2 \pi i} \int \int 
\frac{d z' \wedge d \bar{z}'}{z' - z} (u \Phi_{2 j} )(z',\bar{z}'), \\ 
\label{hom2}
\Phi_{2j}(z,\bar{z}) & = & \frac{1}{2 \pi i} \int \int 
\frac{d z' \wedge d \bar{z}'}{\bar{z}' - \bar{z}} 
(\bar{u} \Phi_{1j} )(z',\bar{z}') e^{- i k_j(z-z') - 
i \bar{k}_j (\bar{z} - \bar{z}')},
\end{eqnarray}
with the boundary conditions as $|z| \to \infty$, 
\begin{equation}
\label{bound}
{\bf \Phi}_j(z,\bar{z}) \rightarrow \frac{{\bf e}_1}{z}.
\end{equation}
Equivalently, this boundary condition can be written as 
renormalization conditions for Eqs. (\ref{hom1}) and (\ref{hom2}), 
\begin{eqnarray}
\label{norm1} 
\frac{1}{2 \pi i} \int\!\!\int d z \wedge d \bar{z} 
(u \Phi_{2j} )(z,\bar{z}) & = & 1, \\
\label{norm2} 
\frac{1}{2 \pi i} \int\!\!\int d z \wedge d \bar{z} (\bar{u} 
\Phi_{1j} )(z,\bar{z}) e^{i (k_j z + \bar{k}_j \bar{z})} & = & 0.
\end{eqnarray}
The other (degenerate) bound state ${\bf \Phi}_j'(z,\bar{z})$ can be 
expressed in terms of ${\bf \Phi}_j(z,\bar{z})$ using Eq. (\ref{N_2}), 
\begin{equation}
\label{symmetr}
{\bf \Phi}_j'(z,\bar{z}) = \mbox{\boldmath $\sigma$} 
\bar{\bf \Phi}_j(z,\bar{z}) e^{-i(k_j z + \bar{k}_j \bar{z})}.
\end{equation}

The behaviour of the eigenfunction $\mbox{\boldmath 
$\mu$}(z,\bar{z},k,\bar{k})$ near the eigenvalue $k = k_j$ becomes 
complicated due to the fact that the double eigenvalue is embedded 
into the two-dimensional essential spectrum of the Dirac system 
(\ref{Dir1}). We prove the following result. \\

{\bf Proposition 2.1.} For smooth data $b(k,\bar{k}) \in C^1$ at 
$k \neq k_j$, the eigenfunction $\mbox{\boldmath $\mu$}(z,\bar{z},
k,\bar{k})$ has a pole singularity at $k \to k_j$ only if 
\begin{equation}
\label{constraint}
b_0 = \frac{1}{2 \pi} \int\!\!\int dz \wedge d \bar{z} \; \left( 
\bar{z} \bar{u} \Phi_{1j} \right)(z,\bar{z}) \; 
e^{i(k_j z + \bar{k}_j \bar{z})} = 0.
\end{equation}
\\

{\em Proof.} Suppose $\mbox{\boldmath $\mu$}(z,\bar{z},k,\bar{k})$
has a pole singularity at $k = k_j$. Then, it can be shown from 
Eq. (\ref{Dir1}) that the meromorphic continuation of $\mbox{\boldmath 
$\mu$}(z,\bar{z},k,\bar{k})$ is given by the limiting relation, 
\begin{equation}
\label{w}
\lim_{k \to k_j} \left[ \mbox{\boldmath $\mu$} (z,\bar{z},k,\bar{k}) - 
\frac{i {\bf \Phi}_j(z,\bar{z})}{k - k_j} \right] = 
( z + z_j) {\bf \Phi}_j(z,\bar{z}) + c_j {\bf \Phi}_j'(z,\bar{z}),
\end{equation}
where $z_j$, $c_j$ are some constants. Using Eqs. (\ref{d-bar}), 
(\ref{b}), and (\ref{N_2}), we find the differential relation 
for $b(k,\bar{k})$,
$$
\frac{\partial b}{\partial \bar{k}} = \frac{b(k,\bar{k})}{2\pi} 
\int\!\!\int dz \wedge d\bar{z} \left( \bar{u} \bar{\mu}_2 \right)
(z,\bar{z}) - \frac{1}{2 \pi i} \int\!\!\int dz \wedge d \bar{z}
\left( \bar{z} \bar{u} \mu_1 \right)(z,\bar{z}) e^{i(kz + \bar{k}\bar{z})}.
$$
In the limit $k \to k_j$, this equation reduces with the help of 
Eqs. (\ref{norm1}) and (\ref{w}) to the form, 
$$
\frac{\partial b}{\partial \bar{k}} = \frac{b(k,\bar{k})}{\bar{k} - 
\bar{k}_j} - \frac{b_0}{k-k_j},
$$
where $b_0$ is given in Eq. (\ref{constraint}). The reduced equation 
exhibits the limiting behavior of $b(k,\bar{k})$ as $k \to k_j$, 
\begin{equation}
\label{asympt_b}
b(k,\bar{k}) \rightarrow - b_0 \frac{\bar{k} - \bar{k}_j}{k - k_j} 
\ln|\bar{k}-\bar{k}_j|.
\end{equation}
On the other hand, it follows from Eqs. (\ref{N_2}), (\ref{symmetr}), 
and (\ref{w}) that ${\bf N}_{\mu}(z,\bar{z},k,\bar{k})$ has the 
limiting behavior, 
\begin{equation}
\label{asympt_N} 
{\bf N}_{\mu}(z,\bar{z},k,\bar{k}) \rightarrow \frac{- i {\bf 
\Phi}_j'(z,\bar{z})}{\bar{k} - \bar{k}_j}.
\end{equation}
According to Eqs. (\ref{asympt_b}) and (\ref{asympt_N}), the 
right-hand-side of Eq. (\ref{d-bar}) is of order ${\rm O}(b_0 |k-k_j|^{-1} 
\ln|k-k_j|)$ as $k \to k_j$. On the other hand, the left-hand-side of 
Eq. (\ref{d-bar}) must be of order $O(1)$ in the limit $k \to k_j$ 
according to Eq. (\ref{w}). Therefore, the eigenfunction 
$\mbox{\boldmath $\mu$}(z,\bar{z},k,\bar{k})$ has a pole at $k = k_j$ 
only if the constraint $b_0 = 0$ holds. $\Box$
\\

The limiting relation (\ref{w}) was introduced by Arkadiev {\em et al} 
\cite{APP}. However, the authors did not notice that the discrete 
spectrum is supported only by potentials which satisfy the additional 
constraint (\ref{constraint}). In particular, such potentials include 
the multi-lump solutions for which $b(k,\bar{k}) = 0$ everywhere 
in the $k$-plane. 

\subsection{Expansion Formulas for Inverse Scattering}

Combining Eq. (\ref{d-bar}) for the essential spectrum and Eq. (\ref{w}) 
for the discrete spectrum, we reconstruct the eigenfunction 
$\mbox{\boldmath $\mu$}(z,\bar{z},k,\bar{k})$ \cite{FA,APP}, 
\begin{equation}
\label{inverse}
\mbox{\boldmath $\mu$}(z,\bar{z},k,\bar{k}) = {\bf e}_1 + 
\sum_{j = 1}^n \frac{i {\bf \Phi}_j(z,\bar{z})}{k - k_j} + 
\frac{1}{2 \pi i} \int\!\!\int \frac{dk' \wedge d \bar{k}'}{k' - k} 
b(k',\bar{k}') {\bf N}_{\mu}(z,\bar{z},k',\bar{k}'), 
\end{equation}
where $n$ is number of distinct eigenvalues $k_j$ of double multiplicity.
At $k \to k_j$, this system is coupled with the algebraic system for 
the bound states, 
\begin{equation}
\label{inverse1}
(z + z_j) {\bf \Phi}_j(z,\bar{z}) + c_j {\bf \Phi}_j'(z,\bar{z}) 
= {\bf e}_1 + \sum_{l \neq j} \frac{i {\bf \Phi}_l(z,\bar{z})}{k_j - k_l} 
+ \frac{1}{2 \pi i} \int\!\!\int \frac{dk \wedge d \bar{k}}{k - k_j} 
b(k,\bar{k}) {\bf N}_{\mu}(z,\bar{z},k,\bar{k}). 
\end{equation}

Expansion (\ref{inverse}) can be related to the inverse scattering 
transform for the potential $u(z,\bar{z})$ \cite{FA,APP}. It follows 
from Eq. (\ref{Dir1}) that the eigenfunction 
$\mbox{\boldmath $\mu$}(z,\bar{z},k,\bar{k})$  
has the asymptotic expansion as $|k| \to \infty$, 
\begin{equation}
\label{connect}
\mbox{\boldmath $\mu$}(z,\bar{z},k,\bar{k}) = {\bf e}_1 + 
\frac{1}{i k} \mbox{\boldmath $\mu$}_{\infty}(z,\bar{z}) 
+ {\rm O}(|k|^{-2}),
\end{equation}
where $\mu_{2 \infty}(z,\bar{z}) = \bar{u}(z,\bar{z})$ and 
$$
\mu_{1 \infty}(z,\bar{z}) = - \frac{1}{2 \pi i} \int \int 
\frac{d z' \wedge d \bar{z}'}{z' - z} (|u|^2 )(z',\bar{z}'). 
$$
We deduce from Eqs. (\ref{inverse}) and (\ref{connect}) that 
the potential $\bar{u}(z,\bar{z})$ is expressed through 
the eigenfunctions of the Dirac system in the form \cite{FA},  
\begin{equation}
\label{poten}
\bar{u}(z,\bar{z}) = - \sum_{j=1}^n \Phi_{2j}(z,\bar{z}) 
- \frac{1}{2 \pi} \int\!\!\int dk \wedge d \bar{k} \; b(k,\bar{k}) \; 
N_{2 \mu}(z,\bar{z},k,\bar{k}).
\end{equation}

Formulas (\ref{mu1}) to (\ref{poten}) constitute a standard framework 
for the inverse scattering transform of the DSII equation with a new 
relation (\ref{constraint}). The existence and uniqueness of 
solutions of the Fredholm integral equations (\ref{mu1}) and (\ref{mu2}) 
and the $\bar{\partial}$ problem (\ref{d-bar}) and (\ref{inverse}) were 
proved in \cite{BC} and \cite{FS} under the small-norm assumption 
for the potential $u(x,y)$, 
$$
\left( \sup_{(x,y) \in {\bf R}^2} |u|(x,y) \right) \; 
\left( \int\!\!\int |u (x,y)| dx dy \right) < \frac{\pi}{8}.
$$
In this case $n = 0$ and $b(k,\bar{k}) \neq 0$. The nonlinear 
two-dimensional Fourier transform associated to this scheme was 
discussed in Examples 8-10 of Chapter 7.7 of Ref. \cite{AFbook}. 
Indeed, the connection formula (\ref{poten}) implies that there 
is a scalar spectral decomposition of $\bar{u}(z,\bar{z})$ through 
$N_{2 \mu}(z,\bar{z},k,\bar{k})$ for $n = 0$. In order to close 
the decomposition, one could use Eqs. (\ref{b}) and (\ref{N_2}) 
to construct a ``completeness relation'' for the expansion of 
$\delta(z'-z)$ in the form,
$$
\delta(z'-z) = - \frac{1}{2 \pi^2 i} \int\!\!\int dk \wedge d \bar{k}
\bar{N}_{2 \mu}(z',\bar{z}',k,\bar{k}) N_{2 \mu}(z,\bar{z},k,\bar{k}). 
$$ 
However, we show in Proposition 3.4 below that a completeness
theorem for Eq. (\ref{Dir1}) is different and is based on the 
set of eigenfunctions of the adjoint Dirac system. 

\section{Basis for a Scalar Spectral Decomposition}

In this section, we specify the adjoint problem for the Dirac system 
(\ref{Dir1}) and establish orthogonality and completeness relations.

\subsection{The Adjoint System}

The adjoint system for Eq. (\ref{Dir1}) is 
\begin{equation}
\label{adjoint}
\mu^a_{1 z} = i k \mu_1^a - u \mu^a_2, \;\;\;\;
\mu^a_{2 \bar{z}} = \bar{u} \mu^a_1, 
\end{equation}
which provides the balance equation,
\begin{equation}
\label{balance}
i (k' - k) \mu_1^a(k') \mu_2(k) = \frac{\partial}{\partial z} 
\left[ \mu_1^a(k') \mu_2(k) \right] - \frac{\partial}{\partial \bar{z}} 
\left[ \mu_2^a(k') \mu_1(k) \right].
\end{equation}
The system (\ref{adjoint}) admits plane solutions 
$\mbox{\boldmath $\mu$}^a(z,\bar{z},k,\bar{k})$ 
and oscillatory-type solutions ${\bf N}_{\mu}^a(z,\bar{z},k,\bar{k})$ 
with the boundary conditions, 
\begin{eqnarray}
\label{b1}
\lim_{|k| \to \infty} \mbox{\boldmath $\mu$}^a(z,\bar{z},k,\bar{k}) 
& = & {\bf e}_2, \\
\label{b2}
\lim_{|k| \to \infty} {\bf N}_{\mu}^a(z,\bar{z},k,\bar{k}) 
e^{-i(kz+\bar{k}\bar{z})} & = & {\bf e}_1.
\end{eqnarray}
The adjoint eigenfunctions ${\bf N}^a_{\mu}(z,\bar{z},k,\bar{k})$ 
can be expressed through the Green functions, 
\begin{eqnarray}
\label{Nmu1a} 
N^a_{1 \mu}(z,\bar{z}) & = & e^{i(kz + \bar{k} \bar{z})} - 
\frac{1}{2 \pi i} \int\!\!\int \frac{d z' \wedge d \bar{z}'}{\bar{z}' 
- \bar{z}} (u N^a_{2 \mu})(z',\bar{z}') 
e^{ik(z-z')+i\bar{k}(\bar{z}-\bar{z}')}, \\ 
\label{Nmu2a} N^a_{2 \mu}(z,\bar{z}) & = & \frac{1}{2 \pi i} 
\int\!\!\int \frac{d z' \wedge d \bar{z}'}{z' - z} 
(\bar{u} N^a_{1 \mu} )(z',\bar{z}'). 
\end{eqnarray}
They are related to the adjoint eigenfunctions $\mbox{\boldmath $\mu$}^a
(z,\bar{z},k,\bar{k})$ by the formula,
\begin{equation}
\label{adjoint_relation}
{\bf N}_{\mu}^a(z,\bar{z},k,\bar{k}) = - \mbox{\boldmath $\sigma$} 
\bar{\mbox{\boldmath $\mu$}}^a(z,\bar{z},k,\bar{k}) 
e^{i(kz + \bar{k} \bar{z})}.
\end{equation}
Using this representation, we prove the following result. 
\\

{\bf Lemma 3.1.} The spectral data $b(k,\bar{k})$ is expressed 
in terms of the adjoint eigenfunctions as
\begin{equation}
\label{b-2}
b(k,\bar{k}) = \frac{1}{2 \pi} \int\!\!\int dz \wedge d \bar{z} 
( \bar{u} N^a_{1 \mu} )(z,\bar{z}).
\end{equation}

{\em Proof.} Multiplying Eq. (\ref{Nmu1a}) by $\bar{u} \mu_1(k)$, 
integrating over $dz \wedge d\bar{z}$ and using Eq. (\ref{mu2}), 
we express $b(k,\bar{k})$ defined in Eq. (\ref{b}) in the form,
\begin{equation}
\label{b_relation}
b(k,\bar{k}) = \frac{1}{2 \pi} \int\!\!\int dz \wedge d \bar{z} \left[ 
\bar{u} \mu_1(k) N^a_{1 \mu}(k) - u \mu_2(k) N^a_{2 \mu}(k) \right]. 
\end{equation}
On the other hand, multiplying Eq. (\ref{mu1}) by $u N_{1\mu}^a(k)$, 
integrating over $dz \wedge d \bar{z}$, and using Eqs. (\ref{Nmu2a}) 
and (\ref{b_relation}), we get Eq. (\ref{b-2}). $\Box$
\\

Suppose now that $k = k_j$ is an isolated double eigenvalue of Eq. 
(\ref{Dir1}) with the bound states ${\bf \Phi}_j(z,\bar{z})$ and 
${\bf \Phi}'_j(z,\bar{z})$ given by Eqs. (\ref{hom1}) -- (\ref{symmetr}). 
Suppose also that $k = k_j^a$ is an eigenvalue of the adjoint system 
(\ref{adjoint}) with the adjoint bound states ${\bf \Phi}_j^a(z,\bar{z})$ 
and ${\bf \Phi}_j^{a \prime}(z,\bar{z})$.  
\\

{\bf Lemma 3.2.} If $k_j$ is a double eigenvalue of the Dirac system 
(\ref{Dir1}), then $k_j$ is also a double eigenvalue of the adjoint 
system (\ref{adjoint}).

{\em Proof.} We use Eq. (\ref{balance}) with $\mbox{\boldmath $\mu$} 
= {\bf \Phi}_j(z,\bar{z})$ and $\mbox{\boldmath $\mu$}^a = 
{\bf \Phi}_j^a(z,\bar{z})$ at $k = k_j$ and $k' = k_j^a$ and 
integrate over $dz \wedge d \bar{z}$ with the help of Eq. (A.3) of 
Appendix A. The contour contribution of the integral vanishes due to 
the boundary conditions (\ref{bound}) and (\ref{bound-2}) and the 
resulting expression is 
$$
(k_j^a - k_j) \int\!\!\int dz \wedge d\bar{z} \;
(\Phi_{1j}^a \Phi_{2j})(z,\bar{z}) = 0.
$$
The relation $k_j^a = k_j$ follows from this formula if the integral is 
non-zero at $k_j^a = k_j$ (which is proved below in Eq. (\ref{orth3})).
The other possibility is when $k_j^a \neq k_j$ but $\Phi_{1j}^a$ is 
orthogonal to $\Phi_{2j}$. We do not consider such a non-generic situation. 
The other bound state ${\bf \Phi}_j^{a \prime}$ at $k_j^a = k_j$ can be 
defined  using the symmetry relation (see Eq. (\ref{sym_adjoint}) below). 
$\Box$
\\

The adjoint bound state ${\bf \Phi}^a_j(z,\bar{z})$ solves the 
homogeneous equations, 
\begin{eqnarray}
\label{bou1} 
\Phi^a_{1 j}(z,\bar{z}) & = & - \frac{1}{2 \pi i} \int\!\!\int 
\frac{d z' \wedge d \bar{z}'}{\bar{z}' - \bar{z}} (u 
\Phi^a_{2 j})(z',\bar{z}') e^{ik_j(z-z')+i\bar{k}_j(\bar{z}-\bar{z}')}, \\ 
\label{bou2} \Phi^a_{2 j}(z,\bar{z}) & = & \frac{1}{2 \pi i} \int\!\!\int 
\frac{d z' \wedge d \bar{z}'}{z' - z} (\bar{u} \Phi^a_{1 j} )(z',\bar{z}')
\end{eqnarray}
with the boundary condition as $|z| \to \infty$, 
\begin{equation}
\label{bound-2}
{\bf \Phi}^a_j(z,\bar{z}) \rightarrow \frac{{\bf e}_2}{z}, 
\end{equation}
and the normalization conditions, 
\begin{eqnarray}
\label{norm-1} 
-\frac{1}{2 \pi i} \int \int d z \wedge d \bar{z} 
(\bar{u} \Phi^a_{1j} )(z,\bar{z}) & = & 1, \\
\label{norm-2} 
\frac{1}{2 \pi i} \int \int d z \wedge d \bar{z} ( u \Phi^a_{2j} 
)(z,\bar{z}) e^{- i (k_j z + \bar{k}_j \bar{z})} & = & 0.
\end{eqnarray}
In addition, the bound state ${\bf \Phi}_j^{a \prime}(z,\bar{z})$
is related to ${\bf \Phi}(z,\bar{z})$ according to the symmetry formula,
\begin{equation}
\label{sym_adjoint}
{\bf \Phi}_j^{a \prime}(z,\bar{z}) = - \mbox{\boldmath $\sigma$} 
\bar{{\bf \Phi}}_j^a(z,\bar{z}) e^{i(k_j z + \bar{k}_j \bar{z})}.
\end{equation}

Using Eqs. (\ref{adjoint})--(\ref{sym_adjoint}), we see that the adjoint 
eigenfunction $\mbox{\boldmath $\mu$}^a(z,\bar{z},k,\bar{k})$ satisfies 
relations similar to those for $\mbox{\boldmath $\mu$}(z,\bar{z},k,\bar{k})$, 
\begin{equation}
\label{d-bar_adjoint}
\frac{\partial \mbox{\boldmath $\mu$}^a}{\partial \bar{k}} = 
- \bar{b}(k,\bar{k}) {\bf N}_{\mu}^a(z,\bar{z},k,\bar{k})
\end{equation}
and 
\begin{equation}
\label{w_adjoint}
\lim_{k \to k_j} \left[ \mbox{\boldmath $\mu$}^a (z,\bar{z},k,\bar{k}) +  
\frac{i {\bf \Phi}_j^a(z,\bar{z})}{k - k_j} \right] = ( z + z_j) 
{\bf \Phi}_j^a(z,\bar{z}) - \bar{c}_j {\bf \Phi}_j^{a \prime}(z,\bar{z}). 
\end{equation}

The expansions for inverse scattering transform of the adjoint 
eigenfunctions can be found in the form, 
\begin{equation}
\label{inverse_adjoint}
\mbox{\boldmath $\mu$}^a(z,\bar{z},k,\bar{k}) = {\bf e}_2 - 
\sum_{j = 1}^n \frac{i {\bf \Phi}_j^a(z,\bar{z})}{k - k_j} -  
\frac{1}{2 \pi i} \int\!\!\int \frac{dk' \wedge d \bar{k}'}{k' - k} 
\bar{b}(k',\bar{k}') {\bf N}_{\mu}^a(z,\bar{z},k',\bar{k}')
\end{equation}
and 
\begin{equation}
\label{inverse1_adjoint}
(z' + z_j) {\bf \Phi}_j^a(z',\bar{z}') - \bar{c}_j {\bf \Phi}_j^{a 
\prime}(z',\bar{z}') = {\bf e}_2 - \sum_{l \neq j} \frac{i 
{\bf \Phi}_l^a(z',\bar{z}')}{k_j - k_l} - \frac{1}{2 \pi i} 
\int\!\!\int \frac{dk \wedge d \bar{k}}{k - k_j} \bar{b}(k,\bar{k}) 
{\bf N}_{\mu}^a(z,\bar{z},k,\bar{k}). 
\end{equation}

\subsection{Orthogonality and Completeness Relations}

Using the Dirac system (\ref{Dir1}) and its adjoint system (\ref{adjoint}), 
we prove the orthogonality and completeness relations for the set of 
eigenfunctions $S = [N_{2 \mu}(k,\bar{k}),\{\Phi_{2 j}\}_{j=1}^n]$ and 
its adjoint set $S^a = [N^a_{1 \mu}(k,\bar{k}),\{\Phi^a_{1 j}\}_{j=1}^n]$.
\\

{\bf Proposition 3.3.} The eigenfunctions $N_{2 \mu}(z,\bar{z},k,\bar{k})$ 
and $\Phi_{2 j}(z,\bar{z})$ are orthogonal to the eigenfunctions 
$N_{1 \mu}^a(z,\bar{z},k,\bar{k})$ and $\Phi_{1 j}^a(z,\bar{z})$ as follows
\begin{equation}
\label{orth1}
\langle N^a_{1 \mu}(k') | N_{2 \mu}(k) \rangle_z = - 2 \pi^2 i \delta(k'-k), 
\end{equation}
\begin{equation}
\label{orth2}
\langle N^a_{1 \mu}(k) | \Phi_{2 j} \rangle_z = 
\langle \Phi^a_{1 j} | N_{2 \mu}(k) \rangle_z = 0, 
\end{equation}
\begin{equation}
\label{orth3}
\langle \Phi^a_{1 l} | \Phi_{2 j} \rangle_z = 2 \pi i \delta_{jl}, 
\end{equation}
where the inner product is defined as
$$
\langle g(k') | f(k) \rangle_z = \int\!\!\int dz \wedge d \bar{z} \;
g(z,\bar{z},k',\bar{k}') \; f(z,\bar{z},k,\bar{k}). 
$$

{\em Proof.} Using Eqs. (\ref{Nmu2}) and (\ref{Nmu1a}), we expand 
the inner product in Eq. (\ref{orth1}) as 
$$
\langle N^a_{1 \mu}(k') | N_{2 \mu}(k) \rangle_z = I_0 
$$
$$
+ \int\!\!\int dz \wedge d \bar{z} \left( u N_{2\mu}^a \right)(z,k') 
e^{-i(k'z+\bar{k}'\bar{z})} I_1(z) - 
\int\!\!\int dz \wedge d \bar{z} \left( \bar{u} N_{1\mu} \right)(z,k) 
e^{i(kz+\bar{k}\bar{z})} I_1(z) 
$$
$$
+ \frac{1}{2 \pi i} \int\!\!\int dz \wedge d \bar{z} 
\left( u N_{2\mu}^a \right)(z,k') e^{-i(k'z+\bar{k}'\bar{z})} 
\int\!\!\int \frac{dz' \wedge d \bar{z}'}{\bar{z}'-\bar{z}} 
\left( \bar{u} N_{1\mu} \right)(z',k) 
e^{i(kz'+\bar{k}\bar{z}')} \left[ I_1(z) - I_1(z') \right],
$$
where   
\begin{equation} \label{I_0}
I_0 = \int\!\!\int dz \wedge d\bar{z} \; 
e^{i(k'-k)z+i(\bar{k}'-\bar{k})\bar{z}} = - 2 \pi^2 i \delta(k'-k)
\end{equation}
and 
\begin{equation} \label{I_1}
I_1(z) = \frac{1}{2 \pi i} \int\!\!\int \frac{dz' \wedge d\bar{z}'}{
\bar{z}'-\bar{z}} e^{i(k'-k)z'+i(\bar{k}'-\bar{k})\bar{z}'} 
= \frac{1}{i(k'-k)} e^{i(k'-k)z+i(\bar{k}'-\bar{k})\bar{z}}.
\end{equation}
The integrals $I_0$ and $I_1(z)$ are computed in Appendix A. Using 
these formulas, we find the inner product in Eq. (\ref{orth1}) in the form, 
$$
\langle N^a_{1 \mu}(k') | N_{2 \mu}(k) \rangle_z = - 2 \pi^2 i 
\delta(k'-k) + \frac{1}{i (k'-k)} {\cal R}(k,k'),
$$
where the residual term ${\cal R}(k,k')$ is expressed in the form, 
$$
{\cal R}(k,k') = \int\!\!\int dz \wedge d \bar{z} \left[ u N_{2\mu}^a (k') 
N_{2\mu}(k) - \bar{u} N_{1\mu}^a(k') N_{1\mu}(k) \right], 
$$
with the help of Eqs. (\ref{Nmu2}) and (\ref{Nmu1a}). We show 
that ${\cal R}(k,k') = 0$ by multiplying Eq. (\ref{Nmu2a}) by 
$\bar{u} N_{1\mu}(k)$, integrating over $dz \wedge d \bar{z}$ 
and using Eq. (\ref{Nmu1}). 

The zero inner products in Eqs. (\ref{orth2}) and (\ref{orth3}) 
for $j \neq l$ are obtained in a similar way with the help of 
the Fredholm's equations for eigenfunctions ${\bf \Phi}_j$, 
${\bf \Phi}_j^a$, ${\bf N}_{\mu}$, and ${\bf N}_{\mu}^a$. 
In order to find the non-zero inner product (\ref{orth3}) for $j = l$ 
we evaluate the following integral by using the same integral equations,
\begin{eqnarray*}
\int\!\!\int dz \wedge d \bar{z} \; \Phi_{1j}^a \mu_2(k) & = & 
\frac{1}{i(k_j-k)} \int\!\!\int dz \wedge d \bar{z} 
\left[ u \Phi_{2j}^a \mu_2(k) - \bar{u} \Phi_{1j}^a \mu_1(k) \right] \\
\phantom{t} & = & \frac{1}{i(k-k_j)} \int\!\!\int dz \wedge d \bar{z} 
\; \bar{u} \Phi_{1j}^a.
\end{eqnarray*}
Using Eq. (\ref{norm-1}), the right-hand-side identifies to 
$\frac{1}{i(k-k_j)}$. Substituting Eq. (\ref{inverse}) in the 
left-hand-side and using the zero inner products (\ref{orth2}) 
and (\ref{orth3}), we find Eq. (\ref{orth3}) for $j = l$. $\Box$ 
\\

{\bf Proposition 3.4.} The eigenfunctions $N_{2 \mu}(z,\bar{z},k,\bar{k})$ 
and $\Phi_{2 j}(z,\bar{z})$ are complete with respect to the adjoint 
eigenfunctions $N_{1 \mu}^a(z,\bar{z},k,\bar{k})$ and 
$\Phi_{1 j}^a(z,\bar{z})$ according to the identity,
\begin{equation}
\label{complet}
\delta(z'-z) = - \frac{1}{2 \pi^2 i} \int\!\!\int dk \wedge d \bar{k} 
N^a_{1 \mu}(z',\bar{z}',k,\bar{k}) N_{2 \mu}(z,\bar{z},k,\bar{k}) 
- \frac{1}{\pi} \sum_{j=1}^n \Phi^a_{1 j}(z',\bar{z}') \Phi_{2 j}(z,\bar{z}).
\end{equation}

{\em Proof.} Using the symmetry relations (\ref{N_2}) and 
(\ref{adjoint_relation}), we express the integral in Eq. (\ref{complet}) as 
\begin{equation}
\label{integral_help}
\langle N^a_{1 \mu}(z') | N_{2 \mu}(z) \rangle_k = \int\!\!\int 
dk \wedge d \bar{k} \; \bar{\mu}^a_2(z',\bar{z}',k,\bar{k}) \bar{\mu}_1 
(z,\bar{z},k,\bar{k}) e^{ik(z'-z) + i\bar{k}(\bar{z}'-\bar{z})}.
\end{equation}
We use Eqs. (\ref{inverse}), (\ref{inverse1}), (\ref{inverse_adjoint}), 
and (\ref{inverse1_adjoint}) and find the pole decomposition for the 
integrand in Eq. (\ref{integral_help}),
$$
\bar{\mu}^a_2 (z') \bar{\mu}_1(z) = 1 + \sum_{j=1}^n \frac{i}{\bar{k} - 
\bar{k}_j} \left[ \bar{c}_j \bar{\Phi}_{2j}^a(z',\bar{z}') 
\bar{\Phi}_{1j}'(z,\bar{z}) + c_j \bar{\Phi}_{2j}^{a \prime}(z',\bar{z}') 
\bar{\Phi}_{1j}(z,\bar{z}) \right] 
$$
$$
+ \sum_{j=1}^n \frac{i}{\bar{k} - \bar{k}_j} (\bar{z} - \bar{z}') 
\bar{\Phi}_{2j}^a(z',\bar{z}') \bar{\Phi}_{1j}(z,\bar{z})   
+ \sum_{j=1}^n \frac{\bar{\Phi}_{2j}^a(z',\bar{z}') 
\bar{\Phi}_{1j}(z,\bar{z})}{(\bar{k} - \bar{k}_j)^2} 
$$
\begin{equation}
\label{integral_help2}
+ \frac{1}{2 \pi i} \int\!\!\int \frac{dk' \wedge d \bar{k}'}{\bar{k}' - 
\bar{k}} \left[ \bar{\mu}_2^a(z') \bar{b} \bar{N}_{1 \mu}(z) - 
\bar{\mu}_1(z) b \bar{N}_{2 \mu}^a(z') \right](k',\bar{k}').
\end{equation}
We substitute (\ref{integral_help2}) into Eq. (\ref{integral_help}) 
and reduce the integral to the form, 
$$
\langle N^a_{1 \mu}(z') | N_{2 \mu}(z) \rangle_k = I_0 - 2 \pi 
\sum_{j=1}^n \left[ \bar{c}_j \bar{\Phi}_{2j}^a(z',\bar{z}') 
\bar{\Phi}_{1j}'(z,\bar{z}) + c_j \bar{\Phi}_{2j}^{a \prime}(z',\bar{z}') 
\bar{\Phi}_{1j}(z,\bar{z}) \right]  \; I_1(k_j) 
$$
$$
+ \sum_{j=1}^n \bar{\Phi}_{2j}^a(z',\bar{z}') \bar{\Phi}_{1j}(z,\bar{z}) 
\left[ 2 \pi (\bar{z}' - \bar{z}) I_1(k_j) + I_2(k_j) \right] 
$$
\begin{equation}
\label{integral_help1}
- \int\!\!\int dk \wedge d \bar{k} I_1(k) \left[  \bar{\mu}_2^a(z') 
\bar{b} \bar{N}_{1 \mu}(z) - \bar{\mu}_1(z) b \bar{N}_{2 \mu}^a(z') 
\right](k,\bar{k}) e^{-ik(z'-z)-i\bar{k}(\bar{z}'-\bar{z})},
\end{equation}
where the integrals $I_0$ and $I_1(k)$ are given in Eqs. (\ref{I_0}) 
and (\ref{I_1}) respectively, with $z$ and $k$ interchanged,
while the integral $I_2(k_j)$ is defined by 
\begin{equation}
\label{I_2}
I_2(k_j) = \lim_{\epsilon \to 0} \int\!\!\int_{|k-k_j|\geq \epsilon} 
\frac{d k \wedge d \bar{k}}{(\bar{k} - \bar{k}_j)^2} e^{ik(z'-z) 
+ i \bar{k} (\bar{z}'-\bar{z})}.
\end{equation}
The integral $I_2(k)$ is found in Appendix A in the form, 
$I_2(k_j) = - 2 \pi (\bar{z}'-\bar{z}) I_1(k_j)$, such that 
the third term in Eq. (\ref{integral_help1}) vanishes. In order 
to express the second term in Eq. (\ref{integral_help1}) we use 
Eqs. (\ref{inverse1}), (\ref{inverse1_adjoint}), (\ref{symmetr}), 
and (\ref{sym_adjoint}) and derive the relation,
$$
- \sum_{j=1}^n \left[ \bar{c}_j \bar{\Phi}_{2j}^a(z',\bar{z}') 
\bar{\Phi}_{1j}'(z,\bar{z}) + c_j \bar{\Phi}_{2j}^{a \prime}(z'\bar{z}') 
\bar{\Phi}_{1j}(z,\bar{z}) \right] e^{ik_j(z'-z) + 
i \bar{k}_j (\bar{z}'-\bar{z})} = 
$$
$$
\sum_{j=1}^n \left[ \bar{c}_j \Phi_{1j}^{a \prime}(z',\bar{z}') 
\Phi_{2j}(z,\bar{z}) + c_j \Phi_{1j}^a(z',\bar{z}') 
\Phi_{2j}'(z,\bar{z}) \right] = 
$$
\begin{equation}
(z'-z) \sum_{j=1}^n \Phi_{1j}^a(z',\bar{z}') \Phi_{2j}(z,\bar{z}) 
+  \frac{1}{2 \pi i} \sum_{j=1}^n \int\!\!\int \frac{dk \wedge 
d \bar{k}}{k - k_j} \left[ \Phi_{2j}(z) \bar{b} N_{1 \mu}^a(z') + 
\Phi_{1j}^a(z') b N_{2 \mu}(z) \right](k,\bar{k}).
\label{help21}
\end{equation}
Using this expression and Eqs. (\ref{I_0}), (\ref{I_1}), and (\ref{I_2}), 
we rewrite Eq. (\ref{integral_help1}) in the form, 
$$
\langle N^a_{1 \mu}(z') | N_{2 \mu}(z) \rangle_k = - 2 \pi^2 i 
\delta(z'-z) - 2 \pi i \sum_{j=1}^n \Phi_{1j}^a(z',\bar{z}') 
\Phi_{2j}(z,\bar{z}) 
$$
$$
+ \frac{i}{z'-z} \int\!\!\int dk \wedge 
d \bar{k} \frac{\partial}{\partial \bar{k}} \left[  \left( \mu_1^a(z') 
+ \sum_{j=1}^n \frac{i \Phi_{1j}^a(z')}{k-k_j} \right) \left( 
\mu_2(z) - \sum_{j=1}^n \frac{i \Phi_{2j}(z)}{k-k_j} \right) \right].
$$
The last integral vanishes according to Eq. (A.3) of Appendix A and 
the boundary conditions (\ref{limit}) and (\ref{b1}). $\Box$ 
\\

Our main result for the spectral decomposition associated to the 
Dirac system (\ref{Dir1}) follows from the above orthogonality and 
completeness relations. \\

{\bf Proposition 3.5.} An arbitrary scalar function $f(z,\bar{z})$
satisfying the condition $f(z,\bar{z}) \sim {\rm O}(|z|^{-2})$ as $|z|
\to \infty$ can be decomposed through the set $S = [ N_{2 \mu}(z,
\bar{z},k,\bar{k}), \{ \Phi_{2 j}(z,\bar{z}) \}_{j=1}^n ]$. \\

{\em Proof.} The spectral decomposition is defined through 
the orthogonality relations (\ref{orth1})--(\ref{orth3}) as 
\begin{equation}
\label{decomp}
f(z,\bar{z}) = \int\!\!\int dk \wedge d\bar{k} \alpha(k,\bar{k}) 
N_{2 \mu}(z,\bar{z},k,\bar{k}) + \sum_{j=1}^n \alpha_j \Phi_{2j}(z,\bar{z}),
\end{equation}
where 
\begin{equation}
\label{coefficients}
\alpha(k,\bar{k}) = - \frac{1}{4 \pi^2} \langle N_{1\mu}^a(k) | f \rangle_z, 
\;\;\;\;\alpha_j = \frac{1}{2 \pi i} \langle \Phi_{1j}^a | f \rangle_z.
\end{equation}
Provided the condition on $f(z,\bar{z})$ is satisfied, we interchange 
integration with respect to $d z \wedge d \bar{z}$ and $d k \wedge 
d \bar{k}$ and use the completeness formula (\ref{complet}). $\Box$ \\

The spectral decomposition presented here is different from that of 
Kiselev \cite{K1,K2}. In the latter approach, the function $f(z,\bar{z})$ 
is spanned by {\em squared} eigenfunctions of the original problem 
(\ref{Dir}) defined according to oscillatory-type behaviour at 
infinity. In our approach, we transformed the system (\ref{Dir}) 
to the form (\ref{Dir1}) and defined the oscillatory-type 
eigenfunctions according to the {\em single} eigenfunctions 
${\bf N}_{\mu}(z,\bar{z},k,\bar{k})$. We also notice that the 
(degenerate) bound states ${\bf \Phi}_j'(z,\bar{z})$ are not relevant 
for the spectral decomposition, although they appear implicitly 
through the meromorphic contributions of the eigenfunctions 
${\bf N}_{\mu}(z,\bar{z},k,\bar{k})$ at $k = k_j$ (see Section 4). 

\section{Perturbation Theory for a Single Lump}

We use the scalar spectral decomposition based on Eq. (\ref{decomp}) 
and develop a perturbation theory for multi-lump solutions of the 
DSII equation. We present formulas in the case of a single lump ($n=1$), 
the case of multi-lump potentials can be obtained by summing 
along the indices $j$, $l$ occuring in the expressions below.

The single-lump potential $u(z,\bar{z})$ has the form \cite{APP}, 
\begin{equation}
\label{single_lump}
u(z,\bar{z}) = \frac{c_j}{|z+z_j|^2 + |c_j|^2} e^{i(k_j z + 
\bar{k}_j \bar{z})}, 
\end{equation}
where $c_j$, $z_j$ are complex parameters. The associated bound states 
follow from Eqs. (\ref{inverse1}) and (\ref{inverse1_adjoint}) as
\begin{eqnarray}
\label{single_bound1}
{\bf \Phi}_j(z,\bar{z}) & = & \frac{1}{|z+z_j|^2 + |c_j|^2} 
\left[ \begin{array}{c} \bar{z} + \bar{z}_j \\ - \bar{c}_j 
e^{-i(k_j z + \bar{k}_j \bar{z})} \end{array} \right], \\
\label{single_bound2}
{\bf \Phi}_j^a(z,\bar{z}) & = & \frac{1}{|z+z_j|^2 + |c_j|^2} 
\left[ \begin{array}{c} c_j e^{i(k_j z + \bar{k}_j \bar{z})} \\ 
\bar{z} + \bar{z}_j \end{array} \right]. 
\end{eqnarray}
We first consider a general perturbation to the single lump
subject to the localization condition, $\Delta u \sim
{\rm O}(|z|^{-2})$ as $|z| \to \infty$. We then derive
explicit formulas for a special form of the perturbation
term $\Delta u(z,\bar{z})$.

\subsection{General Perturbation of a Single Lump}

Suppose the potential is specified as $u^{\epsilon} 
= u(z,\bar{z}) + \epsilon \Delta u (z,\bar{z})$, where $u(z,\bar{z})$ 
is given by Eq. (\ref{single_lump}) and $\Delta u(z,\bar{z})$ is a 
perturbation term. Two bound states ${\bf \Phi}_j(z,\bar{z})$ and 
${\bf \Phi}'_j(z,\bar{z})$ are supported by a single-lump potential 
$u(z,\bar{z})$ at a single point $k = k_j$. The spectral decomposition 
given by Eq. (\ref{decomp}) provides a basis for expansion of 
$\mu^{\epsilon}_2(z,\bar{z},\kappa,\bar{\kappa})$ at $k = \kappa$,
\begin{equation}
\label{mu2-e}
\mu_2^{\epsilon}(z,\bar{z},\kappa,\bar{\kappa}) = \int\!\!\int dk \wedge 
d \bar{k} \; \alpha(k,\bar{k}) N_{2 \mu}(z,\bar{z},k,\bar{k}) + 
\alpha_j \Phi_{2 j}(z,\bar{z})
\end{equation}
where $\alpha(k,\bar{k})$ and $\alpha_j$ are defined by Eq.
(\ref{coefficients}) and depend on the parameter $\kappa$. 
The other component $\mu_1^{\epsilon}(z,\bar{z},\kappa,\bar{\kappa})$ 
can be expressed from Eq. (\ref{Dir1}) as
\begin{equation}
\label{mu1-e}
\mu_1^{\epsilon}(z,\bar{z},\kappa,\bar{\kappa}) = \int\!\!\int 
dk \wedge d \bar{k} \; \alpha(k,\bar{k}) N_{1 \mu}(z,\bar{z},k,\bar{k}) 
+ \alpha_j \Phi_{1 j}(z,\bar{z}) + \epsilon \Delta \mu_1(z,\bar{z}),
\end{equation}
where the remainder term $\Delta \mu_1(z,\bar{z})$ solves the equation,
$$
\left( \Delta \mu_1 \right)_{\bar{z}} = - \Delta u \mu_2^{\epsilon}.
$$
We write the solution of this equation in the form,  
\begin{equation}
\label{solut-mu1}
\Delta \mu_1(z,\bar{z}) = A - \frac{1}{2 \pi i} \int\!\!\int 
\frac{d z' \wedge d \bar{z}'}{z'-z} \left( \Delta u \mu_2^{\epsilon} 
\right)(z',\bar{z}'),
\end{equation}
subject to the boundary condition as $|z| \to \infty$,
$$
\Delta \mu_1(z,\bar{z}) \to A + {\rm O}(z^{-1}),
$$
where $A$ is an arbitrary constant. Using the explicit representation
(\ref{solut-mu1}), we transform Eq. (\ref{Dir1}) into the system of
integral equations for $\alpha(k,\bar{k})$ and $\alpha_j$, 
\begin{equation}
\label{integ1}
\alpha(k,\bar{k}) = \frac{\epsilon}{4 \pi^2 i(k - \kappa)} \left[ 
\int\!\!\int dk' \wedge d\bar{k}' \; K(k,\bar{k},k',\bar{k}') 
\alpha(k',\bar{k}') + K_j(k,\bar{k}) \alpha_j + R(k,\bar{k}) A \right] 
+ {\rm O}(\epsilon^2),
\end{equation}
\begin{equation}
\label{integ2} 
\alpha_j = \frac{\epsilon}{2 \pi (k_j - \kappa)} \left[ \int\!\!\int 
dk \wedge d \bar{k} \; P_j(k,\bar{k}) \alpha(k,\bar{k}) 
+ K_{jl} \alpha_l + R_j A \right] + {\rm O}(\epsilon^2),
\end{equation}
where 
$$
K(k,\bar{k},k',\bar{k}') = \langle {\bf N}^a_{\mu}(k) | 
{\bf N}_{\mu}(k') \rangle_{\Delta u}, \;\;\;\;
K_j(k,\bar{k}) = \langle {\bf N}^a_{\mu}(k) | 
{\bf \Phi}_j \rangle_{\Delta u}, 
$$
$$
P_j(k,\bar{k}) = \langle {\bf \Phi}^a_j | 
{\bf N}_{\mu}(k) \rangle_{\Delta u}, \;\;\;\;
K_{jl} = \langle {\bf \Phi}^a_j | 
{\bf \Phi}_l \rangle_{\Delta u}, \;\;\;\;
$$
and the scalar product for squared eigenfunction is defined as \cite{K1,K2}
$$
\langle {\bf f}(k) | {\bf g}(k') \rangle_h = \int\!\!\int dz \wedge 
d \bar{z} \left[ \bar{h}(z,\bar{z}) f_1(z,\bar{z},k,\bar{k}) 
g_1(z,\bar{z},k',\bar{k}') + h(z,\bar{z}) f_2(z,\bar{z},k,\bar{k}) 
g_2(z,\bar{z},k',\bar{k}') \right]. 
$$
The non-homogeneous terms $R(k,\bar{k})$ and $R_j$ can be computed 
exactly as
\begin{eqnarray*}
R(k,\bar{k}) & = & \int\!\!\int dz \wedge d \bar{z} \; \left( \bar{u} 
N^a_{1 \mu} \right) (z,\bar{z}) = 2 \pi b(k,\bar{k}), \\
R_j & = & \int\!\!\int dz \wedge d \bar{z} \left( \bar{u} \Phi^a_{1 j} 
\right) (z,\bar{z}) = - 2 \pi i,
\end{eqnarray*}
where $b(k,\bar{k}) = 0$ if $n \neq 0$. We solve the system of equations 
(\ref{integ1}) and (\ref{integ2}) asymptotically for $\kappa = k_j + 
\epsilon \Delta \kappa$ and $\Delta \kappa \sim {\rm O}(1)$. The leading 
order behaviour of the integral kernels follows from the asymptotic 
representation (\ref{asympt_N}) as $k \to k_j$,
\begin{equation}
\label{asympt1}
K(k,\bar{k},k',\bar{k}') \to \frac{\bar{K}_{jj}}{(\bar{k} - \bar{k}_j)
(\bar{k}' - \bar{k}_j)}, \;\;\;\; K_j(k,\bar{k}) \to \frac{i P_{jj}}{
\bar{k} - \bar{k}_j}, \;\;\;\;P_j(k,\bar{k}) \to \frac{i \bar{P}_{jj}}{
\bar{k} - \bar{k}_j},
\end{equation}
where 
\begin{equation}
\label{elements}
\bar{K}_{jj} = \langle {\bf \Phi}^{a \prime}_j | 
{\bf \Phi}'_j \rangle_{\Delta u}, \;\;\;\;
P_{jj} = \langle {\bf \Phi}^{a \prime}_j | {\bf \Phi}_j 
\rangle_{\Delta u}, \;\;\;\; \bar{P}_{jj} = - \langle 
{\bf \Phi}^a_j | {\bf \Phi}'_j \rangle_{\Delta u}.
\end{equation}
Here we have used the symmetry constraints (\ref{symmetr}) and 
(\ref{sym_adjoint}). The leading order of $\alpha(k,\bar{k})$ as 
$k \to k_j$ follows from Eq. (\ref{integ1}) as 
$$
\alpha(k,\bar{k}) \to - \frac{\epsilon \Delta \bar{\kappa} 
\beta_j}{2 \pi (k - \kappa) (\bar{k} - \bar{k}_j)},
$$
where $\beta_j$ is not yet defined. We use Eq. (A.4) of Appendix A
to compute the integral term, 
$$
\int\!\!\int \frac{d k \wedge d \bar{k}}{(k - \kappa)(\bar{k} - 
\bar{k}_j)^2} = \frac{2 \pi i}{\bar{k}_j - \bar{\kappa}},
$$
and reduce the system of integral equations (\ref{integ1}) 
and (\ref{integ2}) to an algebraic system as $k \to k_j$,
\begin{eqnarray}
\label{algeb1}
- 2 \pi \Delta \kappa \alpha_j & = & K_{jj} \alpha_j - \bar{P}_{jj} \beta_j 
- 2 \pi i A, \\ \label{algeb2} 2 \pi \Delta \bar{\kappa} \beta_j & = & 
- P_{jj} \alpha_j - \bar{K}_{jj} \beta_j. 
\end{eqnarray}
If $P_{jj} \neq 0$, the determinant of the above system is strictly 
positive. Therefore, homogeneous solutions at $A = 0$ (bound states)
are absent for $\epsilon \neq 0$. This result indicates that the
double eigenvalue at $k = k_j$ disappears under a generic
perturbation of the potential $u(z,\bar{z})$ with $P_{jj} \neq 0$ 
(see also Ref. \cite{GK1}). 

For $A \neq 0$, we find inhomogeneous solutions of Eqs. 
(\ref{algeb1}) and (\ref{algeb2}),
\begin{equation}
\label{solution}
\alpha_j = \frac{2 \pi i A \left( \bar{K}_{jj} + 2 \pi \Delta \bar{\kappa} 
\right)}{|K_{jj} + 2 \pi \Delta \kappa|^2 + |P_{jj}|^2}, \;\;\;\;
\beta_j = \frac{- 2 \pi i A P_{jj}}{|K_{jj} + 2 \pi \Delta \kappa|^2 
+ |P_{jj}|^2}.
\end{equation}
The eigenfunction $\mbox{\boldmath 
$\mu$}^{\epsilon}(z,\bar{z},\kappa,\bar{\kappa})$ 
given by Eqs. (\ref{mu2-e}) and (\ref{mu1-e}) satisfies the boundary 
condition (\ref{limit}) if $A = \epsilon^{-1}$ and has the following 
asymptotic representation, 
\begin{eqnarray}
\nonumber
\mbox{\boldmath $\mu$}^{\epsilon}(z,\bar{z},\kappa,\bar{\kappa}) = {\bf e}_1 
+ \frac{2 \pi i \left[ 2 \pi (\bar{\kappa} - \bar{k}_j) + \epsilon 
\bar{K}_{jj} \right] {\bf \Phi}_j(z,\bar{z})}{|2 \pi (\kappa - k_j) + 
\epsilon K_{jj}|^2 + |\epsilon P_{jj}|^2} & - & \frac{2 \pi i \epsilon 
P_{jj} {\bf \Phi}'_j(z,\bar{z})}{|2 \pi (\kappa - k_j) + \epsilon 
K_{jj}|^2 + |\epsilon P_{jj}|^2} \\
\label{expansion}
& + & \Delta \mbox{\boldmath $\mu$}^{\epsilon}(z,\bar{z}),
\end{eqnarray}
where the term $\Delta \mbox{\boldmath $\mu$}^{\epsilon}(z,\bar{z})$ 
is not singular in the limit $\epsilon \to 0$ and $\kappa \to k_j$. 

In the limit $\epsilon \to 0$, $\kappa \neq k_j$, we find a meromorphic 
expansion for ${\bf \mu}^{\epsilon}(z,\bar{z},\kappa,\bar{\kappa})$ as 
\begin{equation}
\label{expans1}
\mbox{\boldmath $\mu$}^{\epsilon}(z,\bar{z},\kappa,\bar{\kappa}) = 
{\bf e}_1 + \frac{i {\bf \Phi}_j(z,\bar{z})}{\kappa - k_j} + 
\epsilon \left[ \frac{K_{jj} {\bf \Phi}_j(z,\bar{z})}{2 \pi i 
(\kappa - k_j)^2} + \frac{P_{jj} {\bf \Phi}'_j(z,\bar{z})}{2 \pi i 
|\kappa - k_j|^2} \right] + {\rm O}(\epsilon^2).
\end{equation}
It is clear that the double pole can be incorporated by shifting 
the eigenvalue $k_j$ to 
$$
k_j^{\epsilon} = k_j - \frac{\epsilon K_{jj}}{2 \pi}. 
$$
The other double-pole term in the expansion (\ref{expans1}) has a 
non-analytic behaviour in the $k$-plane and leads to the appearance 
of the spectral data $b^{\epsilon}(\kappa,\bar{\kappa}) = \epsilon \Delta 
b(\kappa,\bar{\kappa})$ which measures the departure of $\mbox{\boldmath
$\mu$}^{\epsilon}(z,\bar{z},\kappa,\bar{\kappa})$ from analyticity 
according to Eq. (\ref{d-bar}). We find from Eqs. (\ref{b}) and
(\ref{expans1}) that the spectral data $\Delta b(\kappa,\bar{\kappa})$
has the following singular behaviour as $\kappa \to k_j$
\begin{equation}
\label{DB}
\Delta b(\kappa,\bar{\kappa}) \to \frac{- P_{jj}}{2 \pi |\kappa - k_j|^2}.
\end{equation}
Thus, if $P_{jj} \neq 0$ the analyticity of $\mbox{\boldmath 
$\mu$}^{\epsilon}(z,\bar{z},\kappa,\bar{\kappa})$ is destroyed and the 
lump disappears. This conclusion as well as the analytical solution 
(\ref{solution}) agree with the results of Gadyl'shin and Kiselev 
\cite{GK1,GK2} where the transformation of a single lump into decaying 
wave packet was also studied. 

In the other limit $\epsilon \neq 0$ and $\kappa \to k_j^{\epsilon}$ 
we find another expansion from Eq. (\ref{expansion}), 
\begin{equation}
\label{expans2}
\mbox{\boldmath $\mu$}^{\epsilon}(z,\bar{z},\kappa,\bar{\kappa}) = 
{\bf e}_1 - \frac{2 \pi i}{\epsilon \bar{P}_{jj}} {\bf \Phi}'_j(z,\bar{z}) 
+ {\rm O}(\kappa - k_j^{\epsilon}).
\end{equation}
We conclude that the eigenfunction $\mbox{\boldmath $\mu$}^{\epsilon}(
z,\bar{z},\kappa,\bar{\kappa})$ is now free of pole singularities 
\cite{GK1,GK2}. We summarize the main result in the form of a proposition. 
\\

{\bf Proposition 4.1.} Suppose $u(z,\bar{z})$ is given by Eq. 
(\ref{single_lump}) and $\Delta u(z,\bar{z})$ satisfies the constraint, 
$$
P_{jj} = \langle {\bf \Phi}^{a \prime}_j | {\bf \Phi}_j 
\rangle_{\Delta u} \neq 0.
$$
Then, the potential $u^{\epsilon} = u(z,\bar{z}) + \epsilon \Delta 
u(z,\bar{z})$ does not support embedded eigenvalues of the Dirac system 
(\ref{Dir1}) for $\epsilon \neq 0$. 
\\

\subsection{Explicit Solution for a Particular Perturbation}

Here we specify $c_j = c e^{i \theta}$, where $c$ and $\theta$ are real, 
and consider a particular perturbation $\Delta u(z,\bar{z})$ 
to the lump $u(z,\bar{z})$ (\ref{single_lump}) in the form, 
$$
\Delta u(z,\bar{z}) = Q(z,\bar{z}) e^{i(k_j z + 
\bar{k}_j \bar{z} + \theta )},
$$
where $Q(z,\bar{z})$ is a real function. Using Eqs. (\ref{elements}),
(\ref{single_bound1}), and (\ref{single_bound2}), we find explicitly 
the matrix elements $K_{jj}$ and $P_{jj}$,
$$
K_{jj} = \int\!\!\int dz \wedge d \bar{z} \frac{c (\bar{z} + \bar{z}_j) 
[Q(z,\bar{z}) - \bar{Q}(z,\bar{z})]}{[|z+z_j|^2 + c^2 ]^2} = 0, 
$$
$$
P_{jj} = \int\!\!\int dz \wedge d \bar{z} \frac{|z + z_j|^2 \bar{Q}(z,\bar{z}) 
+  c^2 Q(z,\bar{z})}{[|z+z_j|^2 + c^2 ]^2} = 
\frac{1}{2 c} \int\!\!\int dz \wedge d \bar{z} \left( u \Delta \bar{u} 
+ \bar{u} \Delta u \right). 
$$
The element $P_{jj}$ can be seen as a correction to the field energy, 
$$
N = \frac{i}{2} \int\!\!\int dz \wedge d \bar{z} |u^{\epsilon}|^2(z,\bar{z}) 
= N_0 + i \epsilon c P_{jj} + {\rm O}(\epsilon^2),
$$
where $N_0 = \pi$ is the energy of the single lump solution (independent
of the lump parameters $k_j$ and $c_j$). Thus, a perturbation which leads 
to the destruction of a single lump, that is with $P_{jj} \neq 0$, changes 
necessarily the value for the lump energy $N_0$.  

\section{Concluding Remarks}

The main result of our paper is the prediction of structural instability of 
multi-lump potentials in the Dirac system associated to the DSII equation. 
The multi-lump potentials correspond to eigenvalues embedded into a 
two-dimensional continuous spectrum with the spectral data $b(k,\bar{k})$ 
satisfying the additional constraint (\ref{constraint}). In 
this case, there is no interaction between lumps and continuous radiation. 
However, a generic initial perturbation induces coupling between the lumps 
and radiation and, as a result of their interaction, the embedded 
eigenvalues disappear. This result indicates that the localized multi-lump 
solutions decay into continuous wave packets in the nonlinear dynamics 
of the DSII equation (see also Refs. \cite{GK1,GK2}). 

This scenario is different from the two types of 
bifurcations of embedded eigenvalues discussed in our previous paper 
\cite{PS}. The type I bifurcation arises from the edge of the essential 
spectrum when the limiting bounded (non-localized) eigenfunction is 
transformed into a localized bound state. The type II bifurcation occurs 
when an embedded eigenvalue splits off the essential spectrum. Both 
situations persist in the spectral plane when the essential spectrum 
is one-dimensional and covers either a half-axis or the whole axis. 
However, in the case of the DSII equation, the essential spectrum 
is the whole spectral plane and embedded eigenvalue can not 
split off the essential spectrum. As a result, they disappear 
due to their structural instability.

\section*{Acknowledgements}
We benefited from stimulating discussions with M. Ablowitz, 
A. Fokas, D. Kaup, O. Kiselev, and A. Yurov. D.P. acknowledges 
support from a NATO fellowship provided by NSERC and C.S. 
acknowledges support from NSERC Operating grant OGP0046179. 

\section*{Appendix A. Formulas of the $\partial$-analysis}

Here we reproduce some formulas of the complex $\bar{\partial}$-analysis 
\cite{AFbook} to compute the integrals $I_0$, $I_1(z)$ and $I_2(z)$ 
defined in Eqs. (\ref{I_0}), (\ref{I_1}), and (\ref{I_2}). We define 
the complex integration in the $z$-plane by
$$
\int\!\!\int dz \wedge d \bar{z} f(z,\bar{z}) = 
- \int\!\!\int d \bar{z} \wedge d z f(z,\bar{z}),
$$
where $dz \wedge d\bar{z} = - 2 i dx dy$. The complex $\delta(z)$ 
distribution is defined by  
$$
\int\!\!\int dz \wedge d \bar{z} f(z) \delta(z - z_0) = -2i f(z_0),
\eqno(A.1)
$$
where $\delta(z) = \delta(x) \delta(y)$. In particular, the 
$\delta$-distribution appears in the $\bar{\partial}$-analysis 
according to the relation \cite{AFbook},
$$
\frac{\partial}{\partial \bar{z}} \left[ \frac{1}{z - z_0} \right] 
= \pi \delta(z - z_0).
\eqno(A.2)
$$
Computing the integral $I_0$, we get the formula,
$$
I_0 = \int\!\!\int dz \wedge d \bar{z} e^{ikz + i\bar{k} \bar{z}} =  
- 2 i \int\!\!\int dx dy e^{2 i {\rm Re}(k) x - 2 i {\rm Im}(k) y} = 
- 2 \pi^2 i \delta(k), 
$$
which proves the identity (\ref{I_0}). 

Using the Green's theorem \cite{AFbook}, one has the integration identity, 
$$
\int\!\!\int_D dz \wedge d \bar{z} \left( \frac{\partial f_1}{\partial z} 
- \frac{\partial f_2}{\partial \bar{z}} \right) = \int_C 
\left( f_1 d\bar{z} + f_2 d z \right),
\eqno(A.3)
$$
where $D$ is a domain of the complex plane and $C$ its boundary. The 
generalized Cauchy's formula has the form \cite{AFbook},
$$
f(z,\bar{z}) = \frac{1}{2 \pi i} \int_C \frac{f(z',\bar{z}') d z'}{z' - z} 
+ \frac{1}{2 \pi i} \int\!\!\int_D \frac{dz' \wedge d \bar{z}'}{z'-z} 
\frac{\partial f}{\partial \bar{z}'},
\eqno(A.4)
$$
or, equivalently,  
$$
f(z,\bar{z}) = - \frac{1}{2 \pi i} \int_C \frac{f(z',\bar{z}') d \bar{z}'}{
\bar{z}' - \bar{z}} + \frac{1}{2 \pi i} \int\!\!\int_D \frac{dz' \wedge 
d \bar{z}'}{\bar{z}'-\bar{z}} \frac{\partial f}{\partial z'}.
\eqno(A.5)
$$

In order to find the integral $I_1(z)$ we use Eq. (A.5) with  
$$
f(z,\bar{z}) = \frac{1}{i k} e^{i(kz + \bar{k} \bar{z})}, \;\;\; k \neq 0
$$
and choose the domain $D$ to be a large ball of radius $R$ (see Eq. 
(\ref{comp_integr})). The boundary value integral vanishes since 
$$
\lim_{R \to \infty} \int_{|z|=R} \frac{d \bar{z}}{\bar{z}} 
e^{i(kz+\bar{k}\bar{z})} = - 2 \pi i \lim_{R \to \infty} 
J_0(2|k|R) = 0, 
\eqno(A.6)
$$
where $J_0(z)$ is the Bessel function. Equation (A.5) for the function 
$f(z,\bar{z})$ then reduces to Eq. (\ref{I_1}).

In order to compute the integral $I_2(z_0)$, we apply Eq. (A.3) with
$f_1 = 0$ and 
$$
f_2(z,\bar{z}) = \frac{1}{\bar{z}-\bar{z}_0} e^{i(kz+\bar{k}\bar{z})}.
$$ 
The domain $D$ is chosen as above. The boundary value integral vanishes again, 
$$
\lim_{R \to \infty} \int_{|z| = R} \frac{d z}{\bar{z}} 
e^{i(kz+\bar{k}\bar{z})} = - 2 \pi i \lim_{R \to \infty} 
J_{-2}(2|k|R) = 0, 
\eqno(A.7)
$$
where $J_{-2}(z)$ is the Bessel function. Equation (A.3) for
the function $f_2(z,\bar{z})$ reduces to Eq. (\ref{I_2}).

\end{document}